\documentclass[sigconf,nonacm]{acmart}

\usepackage{url}
\usepackage{algorithm}
\usepackage{algorithmic}
\usepackage{graphicx}
\usepackage{subfig}
\usepackage{listings}
\usepackage{multirow}
\usepackage{color}
\usepackage{xspace}
\usepackage{wrapfig}
\usepackage{enumitem}
\usepackage{soul}
\usepackage{tablefootnote}
\usepackage[mode=build]{standalone}
\usepackage{fancyhdr}
\usepackage{lipsum}
\usepackage{setspace}
\usepackage{listings}
\usepackage{float}
\usepackage{soul}

\usepackage{tikz}
\usetikzlibrary{patterns}
\usetikzlibrary{positioning, intersections, arrows, shapes.geometric, calc, shapes.callouts, decorations.pathreplacing, decorations.markings, shapes}
\usepackage{pifont}
\usepackage{pgfplots}
\usepgfplotslibrary{statistics}
\usepackage{pgfplotstable}
\pgfplotsset{compat=newest}

\AtBeginDocument{%
  }

\newcommand{\FIG}{Figure\xspace}

\newcommand{\EV}{ParaView\xspace}

\setcopyright{acmlicensed}
\copyrightyear{2018}
\acmYear{2018}
\acmDOI{XXXXXXX.XXXXXXX}
\acmConference[Conference acronym 'XX]{Make sure to enter the correct conference title from your rights confirmation email}{June 03--05, 2018}{Woodstock, NY}
\acmISBN{978-1-4503-XXXX-X/2018/06}

\begin{document}

\title{Empowering Student Debugging in Parallel Programming with Execution Traces and Large Language Models}

\author{God’salvation F. Oguibe}
\email{Godsalvation.Oguibe@utsa.edu}
\orcid{0009-0009-9333-2517}
%\email{webmaster@marysville-ohio.com}
\affiliation{%
  \institution{The University of Texas at San Antonio}
  \city{San Antonio}
  \state{TX}
  \country{USA}
}

\author{Vinodh Kumaran Jayakumar}
\email{rvn028@my.utsa.edu}
\affiliation{%
  \institution{The University of Texas at San Antonio}
  \city{San Antonio}
  \state{TX}
  \country{USA}
}

\author{Tongping Liu}
\email{tongping.cs@gmail.com}
\affiliation{%
  \institution{The University of Massachusetts Amherst}
  \city{Amherst}
  \state{Massachusetts}
  \country{USA}
}

\author{Andrew Lan}
\email{andrewlan@cs.umass.edu}
\affiliation{%
  \institution{The University of Massachusetts Amherst}
  \city{Amherst}
  \state{Massachusetts}
  \country{USA}
}

\author{Wei Wang}
\email{Wei.Wang@utsa.edu}
\orcid{0000-0003-2262-2508}
\affiliation{%
  \institution{The University of Texas at San Antonio}
  \city{San Antonio}
  \state{TX}
  \country{USA}
}


\begin{abstract}
Concurrent programming is a core component of Computer Science curricula, yet remains notoriously difficult for students to master due to its inherent complexity and the nondeterministic nature of concurrency bugs such as deadlocks and race conditions. In this work, we present \EV, an educational tool designed to help students understand, debug, and correct concurrency issues in parallel programs written in C/C++. \EV provides transparent execution recording and visualization to make parallel execution observable and comprehensible. We evaluated \EV through a series of debugging and implementation tasks, with 17 students participating. Results showed a significant improvement in debugging and implementation successes compared to previous course iterations. A student survey confirmed that most participants found \EV helpful. To further support learning outside the classroom, we explored using Large Language Models (LLMs) to analyze concurrency bugs and suggest fixes. While LLMs were highly effective in identifying bugs and explaining execution traces, the correctness of their bug fixes varied, especially for more complex synchronization patterns. Our findings suggest that recording-visualization tools like \EV, complemented by artificial intelligence (AI), can improve teaching and learning of concurrent programming.
\end{abstract}


\maketitle

\section{Introduction}\label{sec:introduction}
%Due to the gradual failure of Dennard Scaling since 2005, the growth of single CPU/core performance and clock frequency has been stalled significantly. Therefore, major chip manufacturers shifted to increase the number of cores in order to obtain continuous performance improvement. As multi-core processors become the norm, corresponding software should be written in a multithreaded/concurrent (interchangeably) fashion in order to take advantage of these hardware cores. 
%Therefore, it is imperative for Computer Science (CS) and Computer Engineering (CE) students to be well-prepared for concurrent programming through their undergraduate education. Multiple courses in the CS and CE curricula, such as Operating Systems, Parallel Computing, or System Programming, introduce the concurrency concepts and practices (e.g., process, threads, IPC, multithreading, synchronization), which are also required by ABET Accreditation~\cite{abet}. 

%Since the decline of Dennard Scaling around 2005, improvements in single-core CPU performance and clock frequency have slowed significantly. In response, major chip manufacturers have shifted toward increasing the number of cores to continue advancing overall system performance. As multi-core processors have become standard, software must be developed in a concurrent or multithreaded manner to fully utilize the available hardware resources.
Since the decline of Dennard Scaling around 2005, the improvement in single-core CPU performance and clock frequency has slowed significantly. To maintain progress in overall system performance, major chip manufacturers shifted toward increasing the number of cores. Consequently, multi-core processors have become standard, and software must now be written in a concurrent or multithreaded manner to fully utilize the available hardware resources.

%Consequently, it is essential for undergraduate students in Computer Science (CS) and Computer Engineering (CE) programs to gain a solid foundation in concurrent programming. Core courses within these curricula—such as Operating Systems, Parallel Computing, and Systems Programming—introduce key concurrency concepts and practices, including processes, threads, inter-process communication (IPC), multithreading, and synchronization. These topics also align with the requirements set forth by ABET accreditation standards~\cite{abet}.
As a result, it is crucial for undergraduate students in Computer Science (CS) programs to develop a strong understanding of concurrent programming. Fundamental courses in these programs -- such as Operating Systems, Parallel Computing, and Systems Programming -- cover essential concurrency topics and techniques, including processes, threads, inter-process communication (IPC), multithreading, and synchronization. These topics also correspond to the competencies outlined in the ABET accreditation criteria~\cite{abet}.

To support students in grasping concurrency concepts and building practical skills in concurrent programming, they are commonly assigned multiple concurrent programming projects. However, based on our extensive experience teaching Systems Programming, Operating Systems, and Parallel Computing, we have observed that many students still struggle with concurrent programming—an issue well-documented in existing literature~\cite{ThreadMentor, Convit, ConEE, 10.5555/1231091.1231117, JThreadSpy, Alpaca}. Students frequently introduce a range of concurrency-related bugs~\cite{10.1145/273133.274305, shene1998design} into their programs, such as race conditions, deadlocks, atomicity violations~\cite{10.1145/1346281.1346323}, and order violations~\cite{10.1145/1346281.1346323}. As prior research has shown~\cite{shene1998design, ThreadMentor}, \textit{multiple factors contribute to the inherent difficulty of concurrent programming}:

%(1) it is simply not a human nature to consider multiple tasks concurrently;
%(2) it is difficult to reproduce concurrency bugs of programs, not to mention the debugging, since different thread interleavings will make executions nondeterministic;  
%(3) it is challenging for instructors (and teaching assistants) to support students' learning, as it is infeasible for the instructor to spend hours to debug a student's program, considering more than 50 students in a course. When students cannot get immediate feedback and support, they may become frustrated and are discouraged to pursue further studies and take on a future career that requires concurrent programming. 

%Since concurrent programming is both important and challenging, this project proposes the first step towards \textbf{a paradigm shift in concurrent programming education}. In particular, we will develop \EV{}, a novel educational tool to revamp existing practices in learning and teaching concurrent programming. 
%\EV{} will help students debug their multithreaded programs, providing the debugging feedback, and visualization the reasoning of bugs.
%Moreover, \EV{} will also enable instructors to better assist their students, actively keeping them in the loop.

(1) Humans are generally not naturally inclined to reason about multiple tasks executing concurrently. As a result, it is challenging for students to mentally visualize the execution flow of a parallel program without external tools or guidance.

%(2) Concurrency bugs are inherently difficult to reproduce, and debugging them is even more challenging due to the nondeterministic nature of thread interleavings. Hence, re-executing a buggy program may not consistently reproduce the same issue. Furthermore, running a parallel program inside a debugger (e.g. GDB), can alter its timing behavior, potentially masking or eliminating the bug entirely, making it impossible to use debugger to examine the execution of a buggy parallel program. Without means to examine a buggy execution, students struggle to investigate and understand the root cause of concurrency errors.
(2) Concurrency bugs are inherently difficult to reproduce due to the nondeterministic nature of thread interleavings, making debugging especially challenging. Re-running the same program may not trigger the bug again, and using a debugger like GDB~\cite{GDB} can alter timing behavior, potentially hiding the issue. Without a means to observe buggy execution, students often struggle to identify and understand the root causes of concurrency errors.

%(3) Supporting students’ learning is demanding for instructors and teaching assistants, as it is impractical to spend hours debugging each student’s program in classes when class size is large. Without timely feedback and assistance, students may become frustrated and discouraged from continuing their studies or pursuing careers involving concurrent programming.
(3) Supporting students in learning concurrency is challenging for instructors and teaching assistants, especially in large classes where it's impractical to spend significant time debugging each student's code. Without timely feedback and guidance, students may become frustrated and lose motivation to continue learning or to pursue careers that involve concurrent programming.

%As a result, for best learning outcome, and to address the above issues, it is important that enough supports are provided so that students can learn to debug and correct their parallel programs by themselves. In this paper, we discuss our first step in this direction. In particularly, we present \EV, a novel educational tool to revamp existing practices in learning and teaching concurrent programming focusing on deadlock bugs. 

%To achieve the best learning outcomes and address the challenges outlined above, it is essential to provide sufficient support that empowers students to debug and correct their parallel programs independently. In this paper, we present our initial effort toward this goal. Specifically, we introduce 
%\textbf{\EV}, a novel educational tool designed to enhance the teaching and learning of concurrent programming, with a particular focus on understanding and resolving deadlock and race-condition bugs.
To maximize learning outcomes and address the challenges above, students need effective support to independently debug and correct their parallel programs. In this paper, we present our initial effort toward this goal: \textbf{\EV}, a novel educational tool designed to aid the teaching and learning of concurrent programming using C/C++, with a focus on helping students understand and resolve deadlock and race-condition bugs.

%The core functionality of \EV is execution recording, where it records the order of the invocations of synchronization primitives, such as mutex lock and barrier wait. The recordings include not only the synchronization invocations, but also the id of the thread making the invocation and a time step. This recorded execution allows the students to examine the execution without the need to re-run the problem or using a debuggers.

The core functionality of \EV is execution recording, which captures the order of synchronization primitive invocations, such as mutex locks and barrier waits, during program execution. Each recorded event includes the details of the synchronization operation, the thread ID of the invoking thread, and a timestamp. This recorded execution trace enables students to analyze the program’s behavior without needing to re-run it or rely on traditional debuggers.

%The recording is achieved by using GNU Linker's support for wrapper function, where every invocation to a synchronization function (e.g., mutex lock) is redirected to our corresponding wrapper function, which records the invocation, and then calls the original synchronization. By using wrapper function, this recording is transparent to the students. That is, they would just code as if the recording and wrapper functions do not exist. Moreover, wrapper functions are generally low overhead and would not alter execution for our use case (student parallel programs). In our experience, we observed nearly no overhead with students' programs.

%The recording mechanism is implemented using the GNU Linker's support for wrapper functions. Each call to a synchronization primitive (e.g., \texttt{pthread\_mutex\_lock}) is redirected to a custom wrapper function that logs the invocation before delegating the call to the original synchronization function. This approach ensures that the recording process is completely transparent to students -- they can write their code as usual, without being aware of the underlying instrumentation. This transparency also allows \EV to be used a drop-in solution to existing shared-memory programming assignments without modification to the assignments. Additionally, wrapper functions introduce minimal overhead and do not affect program behavior in our context. In practice, we observed negligible performance impact on student-written parallel programs.
The recording mechanism leverages the GNU Linker's support for wrapper functions. Each synchronization call is redirected to a custom wrapper that logs the invocation before calling the original function (e.g., \texttt{pthread\_mutex\_lock}). This design makes the recording process fully transparent -students can write their code normally without noticing the instrumentation. As a result, \EV can be seamlessly integrated into existing shared-memory programming assignments as a drop-in solution without major changes. Moreover, the wrapper functions introduce minimal overhead and, in practice, have no noticeable effect on the performance (and parallel behaviors) of student  programs.

%To further assist the students to examine the execution, a web-based visualization module is provided to illustrate the timeline of each thread during the execution. This visualization makes it easiest for the students to understand parallel execution progress. The execution recording is also immediately available after execution, providing immediate feedback to the students.

To further support students in analyzing program behavior, \EV includes a web-based visualization module that displays the execution timeline of each thread. This visual representation makes it significantly easier for students to understand the flow and interaction of concurrent threads. %Additionally, the execution trace is available immediately after program execution, offering students prompt feedback to aid in debugging and comprehension.

To assess the effectiveness of \EV in improving students’ debugging skills, we designed four programming tasks. Two involved identifying and fixing deadlocks or race conditions, while the other two involved proper use of condition variables and implementing a barrier. Out of 17 students (12 undergraduates and 5 graduates), 14 (11 undergraduates and 3 graduates) successfully completed all tasks. Notably, all 17 students correctly finished the barrier implementation, which is a significant improvement compared to previous course iterations where no students managed to solve this task. A post-assignment survey completed by 13 students showed that 12 (79\%) found \EV helpful in supporting their concurrent programming.

%While the results suggested that execution recording can help many students debugging parallel programs, for some students, there is still a need to directly help them reasoning the cause of parallel bugs and even provide potential fixes. This task has been traditionally very challenging as it involves machine reasoning. However, with the development in coding with Large Language Model (LLMs), this task may be feasible. Hence, we investigated if LLMs could analyze parallel bugs and generate fixes automatically.

While the results indicate that \EV's execution recording and visualization significantly aids many students in debugging parallel programs, some students still struggle to reason about the root causes of concurrency bugs and to identify appropriate fixes on their own. 
%Traditionally, this has been a particularly challenging task, as it requires complex machine reasoning. However, recent advances in code understanding and generation using Large Language Models (LLMs) have opened up new possibilities. Motivated by these developments, 
To help these students learn outside the classroom, we explored using Large Language Models (LLMs) to automatically analyze parallel code and execution traces, and suggest fixes.

%We tested LLM-supported debugging with five programs with various deadlock issues. Our initial results showed that LLMs could successfully identify the causes of deadlocks in simple parallel programs. For certain programs, LLM can detect the causes of bugs using only the code. For other programs, LLMs did need the execution traces to detect the causes. However, the bugfixes recommended by LLMs may not always work. More specifically, the bugfixes recommended can resolve deadlocks. However, the bugfixes may not always follow the program’s original logic. Fixing bugs with conditional variables is also difficult. To obtain correct bug fixes, specially-design prompts are required to guide the LLM. 

We evaluated LLM-assisted debugging on six programs exhibiting various deadlock issues. Our initial results showed that LLMs can successfully identify the causes of deadlocks and race conditions in parallel programs.
In some cases, the LLM was able to detect the issue using only the source code, while in others, execution traces were necessary for accurate diagnosis. However, the quality of the suggested bug fixes varied. While the proposed fixes often resolved the bugs, they did not always preserve the original program logic. In particular, resolving bugs involving condition variables and atomic instructions proved more challenging. Generating correct and context-aware fixes often required carefully crafted, task-specific prompts to guide the LLM effectively.

The contributions of this paper include:

% 1. The execution recording and visualization tool, \EV, which simplifies parallel programming debugging for students and can be used as a drop-in solution to existing shared-memory programming assignments. 

% 2. The evaluation of \EV in real classroom teaching with 17 students, showing the effectiveness of \EV and the students acceptance of \EV. 

% 3. The exploration result of to what extent LLMs can be used to help student reason the causes of concurrency bugs and generate bugfixes.

1. \EV: An Execution Recording and Visualization Tool – A lightweight, drop-in solution that simplifies the debugging of parallel programs for students, which also can be integrated with existing shared-memory programming assignments.

2. Classroom Evaluation with 17 Students – An empirical study demonstrating the effectiveness of \EV in improving student debugging performance and highlighting strong student acceptance and engagement with the tool.

3. Exploring LLMs for Concurrency Bug Support – An initial investigation into how LLMs can assist students in reasoning about concurrency bugs and generating potential fixes, including insights into their capabilities and limitations.

The rest of this paper is organized as follows:
Section~\ref{sec:related} discusses related work;
Section~\ref{sec:design} introduces the design and implementation of \EV;
Section~\ref{sec:evaluation} provides experimental evaluation results;
Section~\ref{sec:discussion} discusses limitations and future work;
and Section~\ref{sec:conclusion} concludes the paper.

\section{Related Work}\label{sec:related}
%There exist a plethora of visualization and simulation tools for introductory programming courses~\cite{UUhistle,sorva2012visual,CoffeeDregs,gries2008principled,gestwicki2005methodology,gallego2004javamod,george2000erosi,george2002using,gondow2010mierucompiler,gries2002frames,guo2013online,helminen2010jype,hertz2013trace,hundhausen2002meta,isohanni2011students,sorva2013review}. Although these tools showed the benefit of visualization in teaching programming, they cannot handle concurrent programs. 
%There also exist games proposed for teaching concurrent programming ~\cite{zhu2020understanding,inggs2017learning,zhu2019programming,valls2017graph,popovic2018application} and studies that provide visualization for classic synchronization problems, such as dining philosophers and reader-writers~\cite{10.1145/3287324.3287467,10.1145/3287324.3293708}. 
%However, these games and classic problem visualizations were designed for learning concurrent concepts, not for visualizing and debugging student programs. Xie et al. studied the effectiveness of visualization styles for different types of synchronization problems~\cite{xie2008evaluating}. Although this study was not on student debugging, its visualization styles partially inspired \EV's design.

Numerous visualization and simulation tools have been developed for introductory programming courses~\cite{UUhistle,sorva2012visual,CoffeeDregs,gries2008principled,gestwicki2005methodology,gallego2004javamod,george2000erosi,george2002using,gondow2010mierucompiler,gries2002frames,guo2013online,helminen2010jype,hertz2013trace,hundhausen2002meta,isohanni2011students,sorva2013review}, demonstrating the benefits of visualization in programming education. However, these tools do not support concurrent programs.

Games and visualizations have also been proposed for teaching concurrency concepts~\cite{zhu2020understanding,inggs2017learning,zhu2019programming,valls2017graph,popovic2018application,10.1145/3287324.3287467,10.1145/3287324.3293708}, particularly through classic synchronization problems like dining philosophers and readers-writers. While educational, these tools are not designed for visualizing or debugging students’ own concurrent programs.

Xie et al.~\cite{xie2008evaluating} evaluated different visualization styles for synchronization problems; although not focused on student debugging, their findings helped inform aspects of \EV's design.

%ThreadMentor provides a new library with customized thread synchronizations that are different from the POSIX standard~\cite{ThreadMentor}, while \EV is fully compatible with POSIX. Moreover, ThreadMentor cannot help understand the concurrency bugs that are caused by missing synchronizations, such as race conditions.   Kim et al.\ proposed a tool for visualizing only deadlocks~\cite{kim2009visualizing}. Kang et al.\ studied visualization for race conditions~\cite{kang2014visualization}. As a comparison, \EV handles both deadlocks and visualizations, enabling it to detect a wider range of concurrent bugs. 

ThreadMentor introduces a custom thread synchronization library that deviates from the POSIX standard~\cite{ThreadMentor}, whereas \EV remains fully POSIX-compliant. Additionally, ThreadMentor does not support debugging race conditions. Other tools have focused on specific issues—Kim et al.\ visualized only deadlocks~\cite{kim2009visualizing}, while Kang et al.\ addressed race conditions~\cite{kang2014visualization}. In contrast, \EV supports visualization and detection of both deadlocks and race conditions, offering broader coverage of concurrency bugs.

%There also exist concurrent program simulators with visualizations. Convit provides a GUI interface that simulates and visualizes the concurrent execution of small programs using execution traces~\cite{Convit}.   ConEE extends Convit~\cite{ConEE} with a code validator to detect race conditions and deadlocks. Both ConEE and Convit use a special programming language, which limits it applicability in today's popular C/C++ based projects. Furthermore, ConEE and Convit are simulators that can only handle very short programs with short execution times and limited concurrent execution paths. Alpaca employs a stateless model checker to identify the reason of failures~\cite{Alpaca}. However, as a testing tool with model checker, it is too complex for college students, as noted by prior work~\cite{ConEE}.
Several tools simulate and visualize concurrent program execution. Convit offers a GUI-based simulator using execution traces~\cite{Convit}, and ConEE extends it with a code validator for detecting race conditions and deadlocks~\cite{ConEE}. However, both rely on a custom programming language, limiting their applicability to modern C/C++ projects. 
%They also support only small programs with short execution times and limited concurrency.
Alpaca uses a stateless model checker to diagnose failures~\cite{Alpaca}, but its complexity makes it unsuitable for undergraduate education, as noted in prior studies~\cite{ConEE}.

%Deadlock Empire provides a set of code snippets and enables students to manually explore different executions~\cite{DeadlockEmpire}.
%VITTI animates some parallel algorithms~\cite{VITTI},  but it is not designed to aid concurrent program development, debugging, or verification, as noted by the authors themselves.
%Progvis is a visualization tool based on the Storm programming language~\cite{Progvis}, with partial support of C/C++. Moreover, Convit/ConEE, Progivs, VITTI, and Deadlock Empire all require students to manually control and/or schedule executions to identify potential bugs. Some tools aim at Java programs. Bi et al.\ proposed a tool that can display and replay the execution of multithreaded Java programs~\cite{10.5555/1231091.1231117}. It uses AspectJ to collect interested data without changing of programs or recompilation. 
%JThreadSpy collects the execution traces and displays the execution with a UML diagram~\cite{JThreadSpy}. 
%These tools cannot handle popular C/C++ based concurrent programming projects. 

Deadlock Empire presents code snippets that allow students to manually explore different execution paths~\cite{DeadlockEmpire}. VITTI animates parallel algorithms~\cite{VITTI}, but it is not intended for concurrent program development, debugging, or verification, as noted by its authors. Progvis provides visualizations using the Storm language with limited C/C++ support~\cite{Progvis}. Furthermore, Convit, ConEE, Progvis, VITTI, and Deadlock Empire all require manual control or scheduling to identify bugs.

Some tools focus on Java concurrency. Bi et al.\ developed a tool that replays multithreaded Java executions using AspectJ to collect data without modifying or recompiling code~\cite{10.5555/1231091.1231117}. JThreadSpy visualizes execution traces as UML diagrams~\cite{JThreadSpy}. However, none of these tools support debugging or visualization for mainstream C/C++ concurrent programming projects. 

%There are also studies investigating using LLM to assist teaching programming~\cite{CodeAid,2025-Raihan-SIGCSE-LLMEduSurvey,2025-Yan-HSSC-LLMProgEdu,CEMR,2024-Shen-SIGCSE}. Majority of these studies were about basic programming, instead of parallel programming. In the only exception, Estévez-Ayres et al. evaluated using ChatGPT and Bard to detect concurrency bugs by sending student code to them. They found both LLMs inadequate for this task~\cite{2024-Estevez-ParallelEduLLM}. However, our results showed the newer LLMs were more capable at LLM debugging. Moreover, our results also show that execution traces were necessary for the LLMs to correctly bug detection, then just student code. We also explored using LLM to recommend fixes for concurrency bugs than just detecting them.

Several studies have investigated using large language models (LLMs) to support programming education~\cite{CodeAid,2025-Raihan-SIGCSE-LLMEduSurvey,2025-Yan-HSSC-LLMProgEdu,CEMR,2024-Shen-SIGCSE}, though most focus on basic programming rather than parallel programming. An exception is the work of Estévez-Ayres et al., which evaluated ChatGPT and Bard for detecting concurrency bugs using only student code. They found them inadequate for this purpose~\cite{2024-Estevez-ParallelEduLLM}. In contrast, our results show that newer LLMs perform significantly better when provided both execution traces and code. In fact execution traces are crucial for complex programs for accurate bug detection. Furthermore, we explored the use of LLMs not only for identifying concurrency bugs but also for recommending fixes.

%Moreover, in this work, we also explore the possibility of automatic diagnosis and automatic bug fix. This allows \EV{} not only visualize the execution process but also provide feedback on the cause of the concurrent bug, the type of the bug, and hints on potential fixes. Existing tools cannot provide such hints.

\section{Design and Implementation of \EV}\label{sec:design}
\subsection{Execution Recording} 
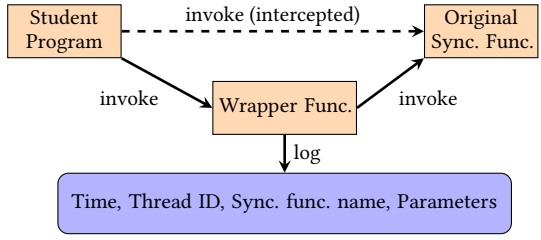
\begin{figure}
    \centering
    \begin{tikzpicture}[font=\fontsize{8}{9}\selectfont, >=stealth,
  scale=1, every node/.style={transform shape}]
  % styles and shapes
  \tikzstyle{io} = [rectangle, rounded corners, minimum width=1cm,
  minimum height=0.8cm, text centered, draw=black, fill=blue!30]
  %\tikzstyle{io} = [trapezium, trapezium left angle=70, trapezium right
  %angle=110, minimum width=1cm, minimum height=0.8cm, text centered, draw=black,
  %fill=blue!30]
  \tikzstyle{process} = [rectangle, minimum width=1.5cm, minimum height=0.7cm, text
  centered, draw=black, fill=orange!30, text width=1.5cm, inner sep = 0]
  \tikzstyle{decision} = [diamond, minimum width=1.6cm, minimum height=0.5cm, text
  centered, draw=black, fill=green!30, inner sep=0]
  \tikzstyle{minilstm} = [draw, minimum height=0.2cm, minimum width=0.2cm,
  fill=YellowGreen!70, inner sep = 0]
  \tikzstyle{arrow} = [->, line width=1]

  % student program
  \node[process, aspect=2.5,  inner sep=0] (studentprog)
  {\fontsize{8}{9}\selectfont  Student Program};

  % synchronization function
  \node[process, right = 4cm of studentprog, aspect=2.5,  inner sep=0] (syncfunc)
  {\fontsize{8}{9}\selectfont  Original Sync. Func.};
  \draw[arrow, dashed] (studentprog.east) -- node[yshift =0.2cm] {invoke (intercepted)} (syncfunc.west);
  
  % wrapper function
  \node[process, below right = 0.3cm and 1.2cm of studentprog, aspect=2.5,  inner sep=0,text width=1.9cm] (wrapper)
  {\fontsize{8}{9}\selectfont  Wrapper Func.};
  \draw[arrow] (studentprog.south east) -- node[yshift = -0.2cm, xshift = -0.5cm] {invoke} (wrapper.west);
  \draw[arrow] (wrapper.east) -- node[yshift = -0.2cm, xshift = 0.5cm] {invoke} (syncfunc.south west);

  % log
  \node[io, below = 0.5cm of wrapper, inner sep=0, text width=6cm] (log)
  {\fontsize{8}{9}\selectfont  Time, Thread ID, Sync. func. name, Parameters};
  \draw[arrow] (wrapper.south) -- node[yshift = 0cm, xshift = 0.3cm] {log} (log.north);

\end{tikzpicture}
    \caption{Intercepting and redirecting synchronization invocations with wrapper functions in \EV.}
    \label{fig:wrapper_arch}
\end{figure}

%As stated previously, the core function of \EV is logging the invocations of synchronization function. This logging is achieved by intercepting student program's calls to synchornization functions (e.g., \texttt{pthread\_mutex\_lock}). 

As previously mentioned, the core functionality of \EV is to log invocations of synchronization functions. This is accomplished by intercepting the student program’s calls to synchronization primitives (e.g., \texttt{pthread\_mutex\_lock}).

%This interception process is illustrated with \FIG~\ref{fig:wrapper_arch}, where the invocation of a original synchronization function was redirected to the wrapper function supplied by \EV. This wrapper logs the details of the call, including timestamp of the call, the ID of the thread making the call, the name of the synchronization function, and the parameters to this function. The parameters are usually the address of the synchronization object and related attributes, such as the address of a Pthread mutex. After logging the invocation, the wrapper calls the original synchronization function to perform the synchronization operation.

This interception mechanism is illustrated in \FIG~\ref{fig:wrapper_arch}, where calls to original synchronization functions are redirected to wrapper functions provided by \EV. Each wrapper logs key details of the invocation, including the timestamp, the ID of the calling thread, the name of the synchronization function, and its parameters. These parameters typically include the address of the synchronization object (e.g., the address of a Pthread mutex) and any related attributes. After logging this information, the wrapper then invokes the original function to carry out the intended synchronization.

%This interception is achieved with GCC linker's wrapper functionality in customized Makefiles. More specifically, during the compilation, we use the \texttt{-wl,--wrapper} option to tell the linker to link calls to all synchronization functions to our wrapper functions first. By utilizing this wrapper functionality, \EV is effectively transparent to the students. That is, the student code still calls the original synchronization functions. These calls are only replaced during compilation. Moreover, this linking-based interception also imposes little overhead.

This interception is implemented using the GCC linker’s wrapper functionality, configured through customized Makefiles. Specifically, during compilation, the \texttt{-Wl,--wrap} option is used to redirect calls to synchronization functions to corresponding wrapper functions. This approach makes \EV fully transparent to the students -- their code continues to call the original synchronization functions, with the redirection occurring only at link time. Additionally, this linking-based interception introduces minimal overhead, without affecting thread execution order.

%Note that, besides this static linking solution, inceptions can also be achieved with dynamic preload, i.e., using \texttt{LD\_PRELOAD} environment variable. Dynamic preload requires no change to Makefiles. Nonetheless, we choose static linking so that students do not need to apply additional environment variables during program executions.

It’s worth noting that, in addition to static linking, interception can also be achieved using dynamic preloading via the \texttt{LD\_PRELOAD} environment variable. This method requires no changes to the Makefile. However, we opted for static linking to avoid requiring students to set additional environment variables during program execution, keeping the experience simpler and more seamless.

%The function name and parameters in the log, can be easiest obtained by examining the call details in the wrapper function. Timestamp is obtained by using CPU's timestamp instruction, \texttt{RDTSC}. For better performance, thread ID is obtained when threads are created and stored as a thread local variable for logging.

The function name and parameters in the log can be easily captured by inspecting the call details within the wrapper function. Timestamps are recorded using the CPU’s timestamp instruction, \texttt{RDTSC}. For efficiency, thread IDs are obtained at thread creation and stored in a thread-local variable for use during logging.

\subsection{Visualization}

\begin{figure*}
    \centering
    \input{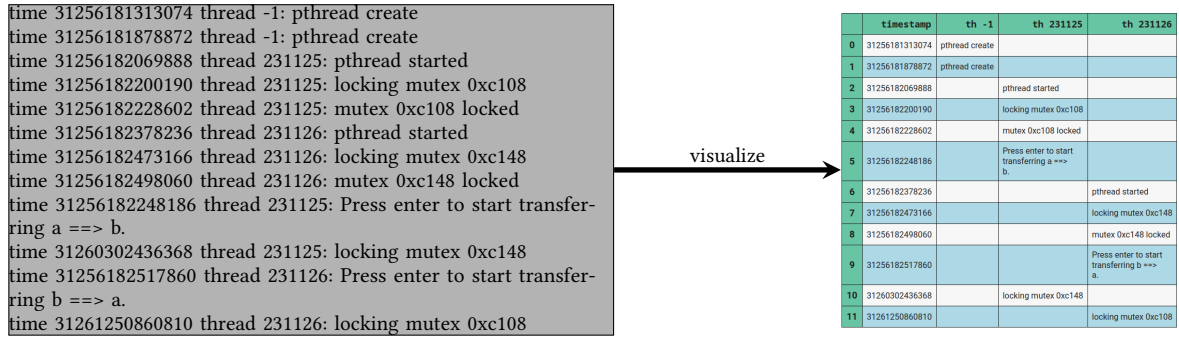}
    \caption{An example of the visualized execution timeline based on recorded logs based on Listing~\ref{lst:practiceone}.}
    \label{fig:visualization_example}
\end{figure*}

%Logs are just a sequences of texts and are difficult to comprehend. As a result, we also provided a web-based visualization module for the logs. As shown with \FIG~\ref{fig:visualization_example}, the visualization module essentially convert the log into a table, where each line represents the operation performance at a specific time in the ascending order. Each column represents a thread. For example, a cell with text "mutex lock 0x888" in the table at row 4 with timestamp "1234" and column with header "thread-5678" means: thread with ID "5678" locked mutex at address 0x888 at time "1234".

Raw logs are plain text sequences that can be difficult to interpret. To address this, we provide a web-based visualization module that transforms the log data into a more intuitive tabular format, as illustrated in \FIG~\ref{fig:visualization_example}. In this view, each row represents an operation performed at a specific point in time (ordered chronologically), while each column corresponds to a thread. For example, a cell containing the text "locking mutex 0x888" in row 4 (timestamp "1234") and under the column labeled "thread-5678" indicates that thread with ID 5678 locked the mutex at address 0x888 at time 1234.

%This visualization allows the student to direct view the execution time of each thread. This visualization module is provide as a web page so that the students do not need to install additional software on the their computer.

This visualization enables students to directly observe the execution timeline of each thread. It is provided as a web-based interface, eliminating the need for students to install any additional software on their computers.

\section{Experimental Evaluation}\label{sec:evaluation}
\subsection{Experiment Setup}

\subsubsection{Methodology}
%We tested \EV in our Parallel Computing class. This class course has 17 students enrolled, including 12 undergraduate students and 5 graduate students. These students participated in two in-class programming practice sessions. The programming practice sessions involved four problems, with two debugging practices and two implementation practices. The students were introduced to \EV at the beginning of these sessions about its usage and purpose. We then guided the student to use \EV to solve the first debugging practice. Once they learned how to use \EV, they proceed to the three practices by themselves with \EV.

We evaluated \EV in our Parallel Computing course, which had 17 students participated -- 12 undergraduate and 5 graduate students. These students participated in two in-class programming practice sessions, which included four problems: two focused on debugging and two on implementation. At the start of the sessions, students were introduced to \EV, including its purpose and usage. We guided them through the first debugging task using \EV. After this initial walkthrough, the students continued independently with the remaining three exercises using \EV to support their work.

\subsubsection{Programming Problems}

\begin{lstlisting}[float, caption={Debugging practice 1 (deadlock, circular wait)} \label{lst:practiceone}]
void* transfer_a_to_b(void *p){
    ...
    pthread_mutex_lock(&lockA);
    account_A_balance -= 100;
    
    pthread_mutex_lock(&lockB);
	account_B_balance += 100;
    pthread_mutex_unlock(&lockB);

    pthread_mutex_unlock(&lockA);
    ...
}
void* transfer_b_to_a(void *p){
    ...
    pthread_mutex_lock(&lockB);
    account_B_balance -= 200;
    
    pthread_mutex_lock(&lockA);
	account_A_ balance += 200;
    pthread_mutex_unlock(&lockA);

    pthread_mutex_unlock(&lockB);
    ...
}
\end{lstlisting}

\begin{lstlisting}[float, caption=Debugging practice 2 (race condition) \label{lst:practicetwo}]
int producer(){
    ...
    pthread_mutex_lock(&mutex);
    count += 2;
    printf("Two product produced\n");
    pthread_cond_signal(&condvar);
    pthread_cond_signal(&condvar);
    pthread_mutex_unlock(&mutex);
    ...
}
int consumer(){
    ...
    pthread_mutex_lock(&mutex);
    pthread_cond_wait(&condvar, &mutex);
    count--;
    printf("consumed one\n");

    pthread_mutex_unlock(&mutex);
    ...
}
\end{lstlisting}

%Listings~\ref{lst:practiceone} and~\ref{lst:practicetwo} gives main part of the code of the two debugging practices. Debugging practice 1 (Listings~\ref{lst:practiceone}) has a deadlock due to a circular waited mutexes -- function \texttt{transfer\_a\_to\_b} in thread 1 locks mutex \texttt{lockA} then mutex \texttt{lockB}, whereas function \texttt{transfer\_a\_to\_b} in thread 2locks these two mutexes in reverse order, causing deadlock. After using \EV to record and visualize the execution, the students could clearly see that these two threads are waiting on each other's lock and entered a deadlock. 

Listings~\ref{lst:practiceone} and~\ref{lst:practicetwo} show the core code segments used in the two debugging practice exercises. Debugging Practice 1 (Listing~\ref{lst:practiceone}) contains a deadlock caused by circular mutex waits: the function \texttt{transfer\_a\_to\_b} in Thread 1 acquires \texttt{lockA} followed by \texttt{lockB}, while the function \texttt{transfer\_b\_to\_a} in Thread 2 acquires the mutexes in the reverse order. This wait leads to a classic deadlock scenario. With \EV to record and visualize the execution, students could clearly observe that the two threads were each waiting for a lock held by the other, leading to deadlock.

%Debugging practice 2 (Listings~\ref{lst:practiceone}) has a race condition, where the \texttt{producer} thread produces two products for the two \texttt{consumer} threads to consume. The  \texttt{producer} and \texttt{consumer} communicates through a condition variable \texttt{condvar}. However, if the \texttt{producer} thread executes two fast, the condition variable signaling may be missed by two \texttt{consumer} threads, causing the race condition. After using \EV to record and visualize the execution, the students could clearly see that \texttt{consumer} threads started waiting on the \texttt{condvar} after it was signaled.
Debugging Practice 2 (Listing~\ref{lst:practicetwo}) involves a race condition between a \texttt{producer} thread and two \texttt{consumer} threads. The \texttt{producer} generates two items, which the \texttt{consumer} threads are meant to consume, with coordination handled via a condition variable, \texttt{condvar}. However, if the \texttt{producer} runs too quickly, its signals to the condition variable may be missed by the consumers, resulting in a race condition. By using \EV to record and visualize the execution, students could clearly observe that the \texttt{consumer} threads began waiting on \texttt{condvar} only after the signal had already been sent, revealing the root cause of the issue.

%The two implementation programs both involves conditional variables, directly following debugging practice 2. In implementation practice 1, the students were asked to implement a chained producer-consumer with three threads: one producer, one consumer-producer, and one consumer. In implementation practice 2, the students were asked to implement a barrier synchronization primitive using conditional variables and mutexes.
%Note that, we have used implementation practice 2 in our past classes, and none of our past students were able to finish the implementation during the in-class session.

\sloppy
Both implementation practices build directly on Debugging Practice 2 and involve the use of condition variables. In Implementation Practice 1, students were tasked with implementing a chained producer-consumer pattern using three threads: a producer, a consumer-producer, and a final consumer. In Implementation Practice 2, students were asked to implement a barrier synchronization primitive using condition variables and mutexes.
It is worth noting that we have used Implementation Practice 2 in previous offerings of the course, and in those instances, no students were able to complete the implementation independently.

\begin{table}[]
\begin{tabular}{|l|l|l|}
\hline
                                & Correct & Incorrect \\ \hline
Debug Practice 1 (deadlock)     & 12/5    & 0/0       \\ \hline
Debug Practice 2 (race cond)    & 11/5    & 1/0       \\ \hline
Implement 1 (chained consumer)  & 11/3    & 1/2       \\ \hline
Implement 2 (barrier implement) & 12/5    & 0/0       \\ \hline
\end{tabular}
\caption{Number of students (undergrad/grad) correctly finished the programming practices.}
\label{tab:evaluation_results}
\end{table}

\subsection{Experiment Results}
%Table~\ref{tab:evaluation_results} gives the numbers of students who correctly finished the four practices using \EV. For the first three practices, all students, including undergraduate and graduate students, were able to correctly finish all of them. Nonetheless, some of them did need assistance in interpreting the outputs of \EV. For the last implementation practice, barrier implementation,all undergraduate students correctly finished it. However, two graduate students were not able to finished it in time. The two graduate students missed part of the practice in-class session due to their research tasks, and hence, were not able to finish this assignment. As a comparison, we had given this barrier implementation before to a previous iteration of the course, and none of the students were able to finish this implementation in class. Therefore, these results of correct implementations show the \EV indeed could help students in parallel program implementation.

Table~\ref{tab:evaluation_results} summarizes the number of students who successfully completed the four programming practices using \EV. 14 students (11 undergraduates and 3 graduates) were able to finish all tasks correctly, though some required guidance in interpreting \EV output initially. Particularly, for the final barrier implementation task, all students completed it successfully. 
For comparison, in a previous offering of the course where the same barrier implementation task was assigned, none of the students completed it during class. These results indicate that \EV has the potential to improve students' ability to implement and debug parallel programs independently. For the cases of failed submissions, it was mostly due to student missing parts of the two classes for this experiment.
%, as they missed part of the in-class session due to research commitments.

\begin{table}
\begin{tabular}{|l|l|l|l|}
\hline
                                     & Yes & No & N/A \\ \hline
Did you use \EV? & 12  & 1  & 0                    \\ \hline
Was \EV helpful?  & 12  & 0  & 1                    \\ \hline
\end{tabular}
\caption{How students responded regarding the usefulness of \EV. 13 Students participated in this survey. "N/A" means "no applicable".}
\label{tbl:survey_results_one}
\end{table}

\begin{table}[]
\begin{tabular}{|l|l|}
\hline
                                                & \# of Responses. \\ \hline
Not necessary, the practices are simple. & 2                    \\ \hline
I don't understand the output/visualization.    & 1                    \\ \hline
The extra visualization step is cumbersome.     & 2                    \\ \hline
I just don't know where to start.               & 1                    \\ \hline
Not applicable to me.                           & 8                    \\ \hline
\end{tabular}
\caption{Students responses to question "if did not use or like \EV, what are the reasons? please select all reasons."}
\label{tbl:survey_results_two}
\end{table}

%After the in-class practice sessions, we also asked the student took a survey regarding the usefulness and design of \EV. 13 students responded to the survey. Table~\ref{tbl:survey_results_one} gives the responses regarding the usefulness. As Table~\ref{tbl:survey_results_one} shows, 12 students have used \EV and found \EV helpful. One student did not use \EV and chose "not applicable" regarding \EV's helpfulness. In the later survey questions, this student mentioned that he thought the practices were simple enough and did not need to use \EV. These surveys results further confirmed that \EV was useful in helping student debugging.

Following the in-class practice sessions, we conducted a survey to gather student feedback on the usefulness and design of \EV. A total of 13 students responded. Table~\ref{tbl:survey_results_one} shows their responses regarding \EV’s usefulness. As shown, 12 students used \EV and reported that it was helpful. One student did not use \EV and marked "not applicable" when asked about its helpfulness. In a later survey response, this student explained that he found the tasks simple and did not feel the need to use \EV. Overall, the survey results further support that \EV was effective in assisting students with parallel programming.

%Table~\ref{tbl:survey_results_two} gives the responses regarding the issues the student faced with using \EV. This is "multiple answers" type question, where student could select any options that applied. As Table~\ref{tbl:survey_results_two} shows, eight students responded with "not applicable", indicating they did not have any issues with \EV. Two students responded that the practices were not difficult enough to justify the use of \EV. One student did not know how to interpret the visualization output, and did not know where to start, due to missing the "walkthrough" part of the class. Two students found the extra visualization is not necessary. These results showed \EV was easy-to-use for most of our students, although students with better programming skills found \EV not necessary, which is expected.

Table~\ref{tbl:survey_results_two} summarizes the student responses regarding any issues they encountered while using \EV. This question was a "select all that apply" question, allowing students to choose multiple relevant options. As shown in the table, eight students selected "not applicable," indicating they did not experience any issues with \EV. Two students felt that the exercises were not difficult enough to warrant using the tool. One student reported difficulty interpreting the visualization output and was unsure how to begin, due to missing the in-class walkthrough. Additionally, two students felt that the extra visualization was unnecessary.
These results suggest that \EV was generally easy to use for most students. Additionally, as expected, students with stronger programming skills were more likely to find the tool unnecessary.

%\EV was not only useful to students. It is worth noting that, implementation a barrier with conditional variable has always been a challenging task in our teaching. For some undergraduate students, we had to provide coding assistance. Nonetheless, \EV had made this assistance process significantly easier. In the past iteration of this barrier practice, we had to teach students how to implement and debug using words and drawing, which still could not clearly convey the core ideas. Moreover, sometimes, students' code was so difficult to understand that we could not determine what when wrong as well. However, with the visualized execution in \EV, it became significantly easier for us to understand student's problem and to explain the proper fix. That is, \EV does not only help student programming, but also makes it easier for the instructor to explain how to program in parallel. 

\EV was not only beneficial to students—it also proved valuable to instructors. Implementing a barrier using conditional variables has consistently been a difficult task in our teaching experience. We had to provide direct coding assistance to some undergraduate students. Traditionally, we relied on verbal explanations and diagrams to help students understand and debug their code, which often failed to clearly convey the core concepts. In some cases, students’ implementations were so convoluted that even we struggled to identify the problem. With \EV’s visualized execution, however, it became significantly easier to pinpoint issues in student code and to clearly explain the appropriate fixes. In this way, \EV not only supports student learning but also enhances the instructor’s ability to teach parallel programming effectively.

\section{Debugging Assistance with LLM}\label{sec:llm}
\begin{table*}[]
\fontsize{8}{9}\selectfont
\begin{tabular}{|l|ll|ll|ll|}
\hline
                       & \multicolumn{2}{c|}{GPT-4o}                      & \multicolumn{2}{c|}{Gemini 2.5 Flash}            & \multicolumn{2}{c|}{Claude 4}                    \\ \hline
                       & \multicolumn{1}{l|}{Can Explain Trace} & Can Provide Fix & \multicolumn{1}{l|}{Can Explain Trace} & Can Provide Fix & \multicolumn{1}{l|}{Can Explain Trace} & Can Provide Fix \\ \hline
Debugging Practice 1       & \multicolumn{1}{l|}{Yes}             & Partially     & \multicolumn{1}{l|}{Yes}             & Partially     & \multicolumn{1}{l|}{Yes}             & Partially     \\ \hline
Debugging Practice 2       & \multicolumn{1}{l|}{Yes}             & Yes (on 2nd try)          & \multicolumn{1}{l|}{Yes}             & Yes           & \multicolumn{1}{l|}{Yes}             & Yes           \\ \hline
Wrong Student Code1    & \multicolumn{1}{l|}{Yes}             & Yes           & \multicolumn{1}{l|}{Yes}             & Yes           & \multicolumn{1}{l|}{Yes}             & Yes           \\ \hline
Wrong Student Code2    & \multicolumn{1}{l|}{Yes}             & Yes           & \multicolumn{1}{l|}{Yes}             & Yes     & \multicolumn{1}{l|}{Yes}             & Yes           \\ \hline
Wrong Student Code3    & \multicolumn{1}{l|}{Yes}             & Yes           & \multicolumn{1}{l|}{Yes}             & Yes           & \multicolumn{1}{l|}{Yes}             & Yes           \\ \hline
Complex Race Condition & \multicolumn{1}{l|}{Partially}             & Partially/No     & \multicolumn{1}{l|}{Yes}             & Partially/No     & \multicolumn{1}{l|}{Yes}             & Partially/No     \\ \hline
\end{tabular}
\caption{Results of using LLM to analyze parallel execution traces and generate code fixes.}
\label{tbl:llm_results}
\vspace{-4mm}
\end{table*}

%As in our teaching, some students still need assistance to understand the visualized log, analyze the cause of bugs, and determine correct bug fixes, we explored automated solutions to provide such assistance so that students can working parallel programming at home. With the recent development in coding with Large Language Models (LLM)~\cite{xxx}, we conducted an preliminary evaluation of using LLMs to reasoning parallel bugs using the code with bug and execution log. In particular, we used three popular LLM services, GPT-4o~\cite{xxx}, Gemini 2.5 Flash~\cite{xxx}, and Claude Sonnet 4~\cite{xxx}. We only used the free version of these these services, so that students do not need to pay for these services if LLMs are truly employed. 

In our teaching, some students still require help interpreting visualized logs, diagnosing bugs, and determining correct fixes. To support independent learning, especially outside the classroom, we explored automated assistance using LLMs. Specifically, we conducted a preliminary evaluation of whether LLMs could analyze parallel bugs using both faulty code and execution logs. We tested three widely available LLMs: GPT-4o~\cite{GPT-4o}, Gemini 2.5 Flash~\cite{Gemini}, and Claude Sonnet 4~\cite{Claude-Sonnet4}. We only used the free version of these these services, so that students do not need to pay for these services if LLMs are truly employed. Table~\ref{tbl:llm_results} summarizes the results of our preliminary LLM study.

%Table~\ref{tbl:llm_results} provides the results of this preliminary study. We started with the debugging practice 1 (Listing~\ref{lst:practiceone}) and debugging practice 2 (Listing~\ref{lst:practiceone}). Both the buggy code of the practices and the execution logs were provided to the LLMs. We then asked the LLMs to explain the logs and code to identify the causes of the concurrency bugs. As Table~\ref{tbl:llm_results} shows, all three LLM services could correctly identify the cause of the bugs. The LLMs could also clear explain the events of execution using the logs and connects the execution events to the code provided to them.
%After identifying the causes, the LLMs were then asked to provide code fixes. For debugging practice 1, all LLMs were able to provide code fixes that eliminates the deadlock. However, the all fixes also incorrectly removed statements accepting user inputs. Hence, the code fix results were marked as "Partially" in Table~\ref{tbl:llm_results}, as the fixes were partially correct. For debugging practice 2, all three LLMs provided the correct fix, although the fixes do differ. GPT-4o and Claude used \texttt{pthread\_cond\_broadcast} to fix the bug, whereas Gemini opted for using two conditional variables. It was also worth noting the GPT-4o gave us wrong bug fix on the first try. However, the following attempts with GPT-4o were all correct, even with new GPT-4o accounts.

We first tested the LLMs on Debugging Practices 1 (Listing~\ref{lst:practiceone}) and 2 (Listings~\ref{lst:practicetwo}), providing both the buggy code and execution logs. Each LLM was asked to interpret the logs and code to identify the root cause of the concurrency bugs. As shown in Table~\ref{tbl:llm_results}, all three LLMs could clearly and correctly explain the events of execution using the logs. Their explanations also correctly linked the execution events to the code provided to them and are easy to understand.

Next, we asked each LLM to suggest code fixes. For Debugging Practice 1, all LLMs proposed fixes that removed the deadlock, but also mistakenly deleted user input statements, so the results were marked as “Partially” correct. For Practice 2, all LLMs provided accurate corrections. GPT-4o and Claude used \texttt{pthread\_cond\_broadcast}, while Gemini used two condition variables. It is also worth noting that GPT-4o initially returned an incorrect fix, but subsequent attempts (even from new accounts) produced correct results.

\begin{lstlisting}[float, caption={An incorrectly submission from a student for debugging practice 2 (red line indicates the bug).}\label{lst:practicetwo_wrongfix}]
int consumer(){
    ...
    pthread_mutex_lock(&mutex);
    @while (count < 2)@ {
  	 pthread_cond_wait(&condvar, &mutex);
    }
    count--;
    printf("consumed one\n");

    pthread_mutex_unlock(&mutex);
    ...
}
\end{lstlisting}

%The debugging practices both involved classical concurrency issues that similar problems may be seen by the LLMs before in their training data. Therefore, we also tested three incorrect parallel programs directly from our students. Listing~\ref{lst:practicetwo_wrongfix} gives one example of incorrect student code. In this example, the student tried to correct debugging practice 2. However, the student code incorrectly used the condition for the while loop on line 4: the condition should be "\texttt{count == 0}" instead of "\texttt{count < 2}". These three incorrect students programs and their execution traces were sent to LLMs for analysis and fixes. As Table~\ref{tbl:llm_results} shows, all three LLMs were able to determine the cause of the bugs and generate correct fixes.

The debugging practices focused on classical concurrency issues, which we may have appeared in the LLMs’ pre-training data. To further assess their capabilities, we also tested three incorrect parallel programs written by students. Listing~\ref{lst:practicetwo_wrongfix} shows one such example, where a student attempted to fix Debugging Practice 2 but made a logic error. Specifically, the student incorrectly used the condition \texttt{count < 2} in the \texttt{while} loop on line 4, when it should be \texttt{count == 0}. 

We provided these three student-written buggy programs and their corresponding execution logs to the LLMs. As shown in Table~\ref{tbl:llm_results}, all three LLMs successfully diagnosed the root causes and generated accurate corrections. 

\begin{lstlisting}[float, caption={A barrier wait implmentation with race condition}\label{lst:barrier_wait}]
typedef struct _my_barrier{
    /* max number of threads using this barrier */
	int max_count; 
    /* current number of threads waiting at barrier */
	int cur_count; 
    /* sequence number of barrier waits */
	int seq; 
} my_barrier_t;

int my_barrier_wait(my_barrier_t *barrier)
    /* get current sequence number */
	int cur_seq = barrier->seq;
	/* increment the counter */
	int old_count = __sync_fetch_and_add(
                        &barrier->cur_count,1);
	if(old_count< (barrier->max_count-1)){
		/* wait if some threads are missing */
		while(cur_seq == barrier->seq){};
	}
	else{ 
		/* reset the counter and increment seq*/
		@barrier->seq++;@
		printf("reset sequence\n");
		@barrier->cur_count = 0;@
		printf("reset count\n");
	}
	return 0;
\end{lstlisting}

%Finally, we tested using a more difficult race condition problem. The code of this problem is given in Listing~\ref{lst:barrier_wait}. The race condition is due to the code from line 22 to line 25, where the resets to \texttt{seq} and \texttt{cur\_count} for a new round of barrier wait are not atomic. That is, after the reset of the \texttt{seq}, there is a chance that another thread calls \texttt{my\_barrier\_wait} and changes the value of \texttt{cur\_count} before it is reset. This race condition is more subtle and difficult to understand. However, the LLMs still performed well on this problem. As Table~\ref{tbl:llm_results} shows, both Gemini and Claude were able to identify the cause of the bug are due to the separated resets. GPT-4o also found out that resets were incorrect. However, it also incorrectly assumed the \texttt{while} loop on line 18 was wrong. 

Finally, we evaluated the LLMs using a more challenging race condition example, shown in Listing~\ref{lst:barrier_wait}. The bug arises from lines 22 to 25, where the updates of \texttt{seq} and \texttt{cur\_count} for the next round of barrier synchronization are not performed atomically. Specifically, after \texttt{seq} is incremented, another thread may enter \texttt{my\_barrier\_wait} and modify \texttt{cur\_count} before it has been reset, leading to a race condition. This issue is subtle and harder to detect. Nevertheless, the LLMs handled it well. As shown in Table~\ref{tbl:llm_results}, both Gemini and Claude correctly identified the non-atomic resets as the root cause using both the execution traces and code. GPT-4o also recognized this problem but also mistakenly flagged the \texttt{while} loop on line 18 as incorrect.

%Correcting the resets in Listing~\ref{lst:barrier_wait}, is more difficult. When asked to correct the code, all three LLMs employed Pthread mutexes and/or conditional variables. While such fixes are correct, they are also slow for a barrier implementation. Hence, we asked the LLMs to fix the bug without using mutexes and conditional variables. Unfortunately, none of the LLMs were able to provide correct fixes this time.

However, correcting the reset logic in Listing~\ref{lst:barrier_wait} is more challenging. When prompted to correct the code, all three LLMs generated fixes using Pthread mutexes and/or condition variables. While these solutions were functionally correct, they introduced unnecessary overhead for a barrier implementation. We then asked the LLMs to resolve the issue without using mutexes or condition variables. However, none of the models could produce a correct fix under this constraint, as the correction involves atomic memory access.

%Overall, our experience showed that LLMs are extremely good at identifying the causes of bugs and explaining the execution logs. The explanation are extremely clear and easy to understand. However, generating bug fixes can still be difficult for current LLMs, which is expected, as correcting concurrency bugs is also challenging even for human.

Overall, we found that LLMs are highly effective at identifying the root causes of concurrency bugs and explaining execution logs when provided both logs and code. Their explanations are typically clear and easy to follow. However, correctly fixing complicated and nuanced bugs may still be challenging for current LLMs. %, especially when bugs become more nuanced and fixes become complicated. 

%—an expected limitation. %, given that resolving concurrency issues is difficult even for experienced programmers.

\section{Discussion and Future Work}\label{sec:discussion}

\textbf{Recording Memory Accesses}.
%Many race condition bugs were due to synchronized memory accesses to variables. Currently, \EV is not capable of recording accesses to variables transparent (i.e., recording these accesses requires explicit calls to logging functions). We plan to add this transparent variable access logging support in the future by adding a compiler plugin or runtime instrumentation to monitor variable accesses. Runtime overhead and monitoring accuracy will be carefully balanced weighting these choices.
Many race condition bugs stem from unsynchronized memory accesses to shared variables. Currently, \EV does not support transparent logging of such accesses. Hence, students must insert explicit logging calls manually. To address this limitation, we will add support for automatic variable access tracking through a compiler plugin or runtime instrumentation. We will carefully balance runtime overhead against monitoring accuracy to ensure it remains practical for education.

\textbf{Supporting other Concurrency Bugs}.
In the future, we plan to expand \EV to support concurrency bugs beyond deadlock and race condition, such as atomicity violation, order violations, priority inversion, and performance issues. We will also design corresponding coding exercises and conduct additional experiments.

\textbf{Integrating \EV with LLM}.
%As LLM shows promises in concurrency education, we will integrate \EV with LLM to evaluate how to better support student learning outside classroom. In particular, there needs additional reason to study how to effectively use LLMs with execution traces and visualization to provide reliably code reasoning and correction suggestions.
As large language models (LLMs) show promise in supporting concurrency education, we plan to integrate \EV with LLMs to further assist student learning beyond the classroom. In particular, further research is needed to understand how to effectively leverage LLMs alongside execution traces and visualizations to provide reliable reasoning about code and accurate bug-fixing suggestions.

\section{Conclusion}\label{sec:conclusion}
Concurrent programming remains a challenging topic for students, often requiring significant support to master. We introduced \EV, a tool that visualizes thread execution and synchronization, helping students better understand and debug concurrency issues. Our classroom deployment showed clear improvements in student performance, with survey responses confirming its usefulness.
To further support independent learning, we explored LLMs for analyzing code and execution traces. While LLMs reliably identified bugs and interpreted logs, generating correct fixes—especially for complex synchronization—was less consistent. Still, the combination of tools like \EV and LLM assistance shows strong potential to enhance the teaching and learning of parallel programming.

\section*{Acknowledgment}
This work was supported by NSF awards 2215359 and 2215193. This paper was edited with the assistance of ChatGPT to help refine the writing and enhance clarity. The content and ideas presented in this paper are entirely the authors' own. 

\newpage
\bibliographystyle{ACM-Reference-Format}
\bibliography{bibfiles/education,bibfiles/parallel_edu}

@inproceedings{ConEE,
author = {Offenwanger, Anna and Lucet, Yves},
title = {ConEE: An Exhaustive Testing Tool to Support Learning Concurrent Programming Synchronization Challenges},
year = {2014},
isbn = {9781450328999},
url = {https://doi.org/10.1145/2597959.2597972},
doi = {10.1145/2597959.2597972},
booktitle = {Proceedings of the Western Canadian Conference on Computing Education},
articleno = {11},
numpages = {6},
location = {Richmond, BC, Canada},
series = {WCCCE '14}
}

@inproceedings{Progvis,
  title={Pilot Study of a Visualization Tool for Object Graphs and Concurrency via Shared Memory},
  author={Str{\"o}mb{\"a}ck, Filip and Mannila, Linda and Kamkar, Mariam},
  booktitle={Proceedings of the 52nd ACM Technical Symposium on Computer Science Education},
  pages={1294--1294},
  year={2021}
}

@misc{DeadlockEmpire,
author={ Petr Hudecek and Michal Pokorny},
title = {The Deadlock Empire},
howpublished = {"\url{https://deadlockempire.github.io/}"},
year={2021}
}

@inproceedings{Convit,
  title={Convit, a tool for learning concurrent programming},
  author={J{\"a}rvinen, Hannu-Matti and Tiusanen, Mikko and Virtanen, Antti},
  booktitle={E-Learn: World Conference on E-Learning in Corporate, Government, Healthcare, and Higher Education},
  pages={2220--2223},
  year={2003},
  organization={Association for the Advancement of Computing in Education (AACE)}
}

@article{10.5555/1231091.1231117,
author = {Bi, Yaodong and Beidler, John},
title = {A Visual Tool for Teaching Multithreading in Java},
year = {2007},
issue_date = {June 2007},
publisher = {Consortium for Computing Sciences in Colleges},
address = {Evansville, IN, USA},
volume = {22},
number = {6},
issn = {1937-4771},
journal = {Journal of Computing Sciences in Colleges},
month = jun,
pages = {156–163},
}

@article{ThreadMentor,
author = {Carr, Steve and Mayo, Jean and Shene, Ching-Kuang},
title = {ThreadMentor: A Pedagogical Tool for Multithreaded Programming},
year = {2003},
issue_date = {March 2003},
volume = {3},
number = {1},
issn = {1531-4278},
url = {https://doi.org/10.1145/958795.958796},
doi = {10.1145/958795.958796},
journal = {Journal on Educational Resources in Computing},
month = mar,
pages = {1–es},
numpages = {30}
}

@INPROCEEDINGS{JThreadSpy,
  author={Malnati, Giovanni and Cuva, Caterina Maria and Barberis, Claudia},
  booktitle={2008 International Conference on Computer Science and Software Engineering}, title={JThreadSpy: A Tool for Improving the Effectiveness of Concurrent System Teaching and Learning}, 
  year={2008},
  volume={5},
  pages={549-552},
  doi={10.1109/CSSE.2008.11}
}

@inproceedings{Alpaca,
author = {Sadowski, Caitlin and Ball, Thomas and Bishop, Judith and Burckhardt, Sebastian and Gopalakrishnan, Ganesh and Mayo, Joseph and Musuvathi, Madanlal and Qadeer, Shaz and Toub, Stephen},
title = {Practical Parallel and Concurrent Programming},
year = {2011},
isbn = {9781450305006},
url = {https://doi.org/10.1145/1953163.1953222},
doi = {10.1145/1953163.1953222},
booktitle = {Proceedings of the 42nd ACM Technical Symposium on Computer Science Education},
pages = {189–194},
numpages = {6},
location = {Dallas, TX, USA},
series = {SIGCSE '11}
}

@misc{abet, 
author = {{ABET organization}},
title = {Criteria for Accrediting Computing Programs, 2021 – 2022},
howpublished = {"\url{https://www.abet.org/accreditation/accreditation-criteria/criteria-for-accrediting-computing-programs-2021-2022/}"},
year={2021}
}

@article{shene1998design,
  title={The design of a multithreaded programming course and its accompanying software tools},
  author={Shene, Ching-Kuang and Carr, Steve},
  journal={The Journal of Computing in Small Colleges},
  volume={14},
  number={1},
  pages={12--24},
  year={1998}
}

@inproceedings{10.1145/273133.274305,
author = {Shene, Chin-Kuang},
title = {Multithreaded Programming in an Introduction to Operating Systems Course},
year = {1998},
isbn = {0897919947},
publisher = {Association for Computing Machinery},
address = {New York, NY, USA},
url = {https://doi.org/10.1145/273133.274305},
doi = {10.1145/273133.274305},
booktitle = {Proceedings of the Twenty-Ninth SIGCSE Technical Symposium on Computer Science Education},
pages = {242–246},
numpages = {5},
location = {Atlanta, Georgia, USA},
series = {SIGCSE '98}
}

@inproceedings{10.1145/1346281.1346323,
author = {Lu, Shan and Park, Soyeon and Seo, Eunsoo and Zhou, Yuanyuan},
title = {Learning from Mistakes: A Comprehensive Study on Real World Concurrency Bug Characteristics},
year = {2008},
isbn = {9781595939586},
publisher = {Association for Computing Machinery},
address = {New York, NY, USA},
url = {https://doi.org/10.1145/1346281.1346323},
doi = {10.1145/1346281.1346323},
booktitle = {Proceedings of the 13th International Conference on Architectural Support for Programming Languages and Operating Systems},
pages = {329–339},
numpages = {11},
location = {Seattle, WA, USA},
series = {ASPLOS XIII}
}

@inproceedings{UUhistle,
author = {Sorva, Juha and Sirki\"{a}, Teemu},
title = {UUhistle: A Software Tool for Visual Program Simulation},
year = {2010},
booktitle = {Proceedings of the 10th Koli Calling International Conference on Computing Education Research},
}

@book{sorva2012visual,
  title={Visual program simulation in introductory programming education},
  author={Sorva, Juha},
  year={2012},
  publisher={Aalto University}
}

@inproceedings{VITTI,
author = {Manickam, Viswanathan and Aravind, Alex},
title = {If a Picture is Worth a Thousand Words, What Would an Animation Be Worth?},
year = {2011},
booktitle = {Proceedings of the 16th Western Canadian Conference on Computing Education},
}

@inproceedings{CoffeeDregs,
  title={Visualization of Object-oriented (Java) Programs.},
  author={Huizing, Cornelis and Kuiper, Ruurd and Luijten, Christian and Vandalon, Vincent},
  booktitle={CSEDU (1)},
  pages={65--72},
  year={2012}
}

@inproceedings{gries2008principled,
  title={A principled approach to teaching OO first},
  author={Gries, David},
  booktitle={Proceedings of the 39th SIGCSE technical symposium on Computer science education},
  pages={31--35},
  year={2008}
}

@inproceedings{gestwicki2005methodology,
  title={Methodology and architecture of JIVE},
  author={Gestwicki, Paul and Jayaraman, Bharat},
  booktitle={ACM symposium on Software visualization},
  pages={95--104},
  year={2005}
}

@inproceedings{gallego2004javamod,
  title={JavaMod: An integrated Java model for Java software visualization},
  author={Gallego-Carrillo, Micael and Gort{\'a}zar-Bellas, Francisco and Vel{\'a}zquez-Iturbide, J Angel},
  booktitle={Program Visualization Workshop},
  pages={102},
  year={2004}
}

@article{george2000erosi,
  title={EROSI—visualising recursion and discovering new errors},
  author={George, Carlisle E},
  journal={ACM SIGCSE Bulletin},
  volume={32},
  number={1},
  pages={305--309},
  year={2000},
  publisher={ACM New York, NY, USA}
}

@article{george2002using,
  title={Using visualization to aid program construction tasks},
  author={George, Carlisle E},
  journal={ACM SIGCSE Bulletin},
  volume={34},
  number={1},
  pages={191--195},
  year={2002},
  publisher={ACM New York, NY, USA}
}

@inproceedings{gondow2010mierucompiler,
  title={Mierucompiler: integrated visualization tool with" horizontal slicing" for educational compilers},
  author={Gondow, Katsuhiko and Fukuyasu, Naoki and Arahori, Yoshitaka},
  booktitle={ACM technical symposium on Computer science education},
  pages={7--11},
  year={2010}
}

@article{gries2002frames,
  title={Frames and folders: A teachable memory model for Java},
  author={Gries, Paul and Gries, David},
  journal={Journal of Computing Sciences in Colleges},
  volume={17},
  number={6},
  pages={182--196},
  year={2002},
  publisher={Citeseer}
}

@inproceedings{guo2013online,
  title={Online python tutor: embeddable web-based program visualization for cs education},
  author={Guo, Philip J},
  booktitle={Proceeding of the 44th ACM technical symposium on Computer science education},
  pages={579--584},
  year={2013}
}

@inproceedings{helminen2010jype,
  title={Jype-a program visualization and programming exercise tool for Python},
  author={Helminen, Juha and Malmi, Lauri},
  booktitle={Proceedings of the 5th international symposium on Software visualization},
  pages={153--162},
  year={2010}
}

@inproceedings{hertz2013trace,
  title={Trace-based teaching in early programming courses},
  author={Hertz, Matthew and Jump, Maria},
  booktitle={Proceeding of the 44th ACM technical symposium on Computer science education},
  pages={561--566},
  year={2013}
}

@article{hundhausen2002meta,
  title={A meta-study of algorithm visualization effectiveness},
  author={Hundhausen, Christopher D and Douglas, Sarah A and Stasko, John T},
  journal={Journal of Visual Languages \& Computing},
  volume={13},
  number={3},
  pages={259--290},
  year={2002},
  publisher={Elsevier}
}

@inproceedings{isohanni2011students,
  title={Students' long-term engagement with the visualization tool VIP},
  author={Isohanni, Essi and Knobelsdorf, Maria},
  booktitle={Proceedings of the 11th Koli Calling International Conference on Computing Education Research},
  pages={33--38},
  year={2011}
}

@article{sorva2013review,
  title={A review of generic program visualization systems for introductory programming education},
  author={Sorva, Juha and Karavirta, Ville and Malmi, Lauri},
  journal={ACM Transactions on Computing Education (TOCE)},
  volume={13},
  number={4},
  pages={1--64},
  year={2013},
  publisher={ACM New York, NY, USA}
}

@misc{zhu2020understanding,
      title={Understanding Learners' Problem-Solving Strategies in Concurrent and Parallel Programming: A Game-Based Approach}, 
      author={Jichen Zhu and Katelyn Alderfer and Brian Smith and Bruce Char and Santiago Ontañón},
      year={2020},
      eprint={2005.04789},
      archivePrefix={arXiv},
      primaryClass={cs.HC}
}

@inproceedings{inggs2017learning,
  title={Learning Concurrency Concepts while Playing Games.},
  author={Inggs, Cornelia P and Gadd, Taun and Giffard, Justin},
  booktitle={CSEDU (1)},
  pages={597--602},
  year={2017}
}

@inproceedings{zhu2019programming,
  title={Programming in game space: how to represent parallel programming concepts in an educational game},
  author={Zhu, Jichen and Alderfer, Katelyn and Furqan, Anushay and Nebolsky, Jessica and Char, Bruce and Smith, Brian and Villareale, Jennifer and Onta{\~n}{\'o}n, Santiago},
  booktitle={Proceedings of the 14th International Conference on the Foundations of Digital Games},
  pages={1--10},
  year={2019}
}

@inproceedings{valls2017graph,
  title={Graph grammar-based controllable generation of puzzles for a learning game about parallel programming},
  author={Valls-Vargas, Josep and Zhu, Jichen and Onta{\~n}{\'o}n, Santiago},
  booktitle={Proceedings of the 12th International Conference on the Foundations of Digital Games},
  pages={1--10},
  year={2017}
}

@inproceedings{kim2009visualizing,
  title={Visualizing potential deadlocks in multithreaded programs},
  author={Kim, Byung-Chul and Jun, Sang-Woo and Hwang, Dae Joon and Jun, Yong-Kee},
  booktitle={International Conference on Parallel Computing Technologies},
  pages={321--330},
  year={2009},
  organization={Springer}
}

@article{popovic2018application,
  title={Application of social game context to teaching mutual exclusion},
  author={Popovi{\'c}, Miroslav and Vladimir, Klemo and {\v{S}}ili{\'c}, Marin},
  journal={Automatika: {\v{c}}asopis za automatiku, mjerenje, elektroniku, ra{\v{c}}unarstvo i komunikacije},
  volume={59},
  number={2},
  pages={208--219},
  year={2018},
  publisher={KoREMA-Hrvatsko dru{\v{s}}tvo za komunikacije, ra{\v{c}}unarstvo, elektroniku, mjerenja~…}
}

@inproceedings{10.1145/3287324.3287467,
author = {Adams, Joel C. and Koning, Elizabeth R. and Hazlett, Christiaan D.},
title = {Visualizing Classic Synchronization Problems: Dining Philosophers, Producers-Consumers, and Readers-Writers},
year = {2019},
booktitle = {Proceedings of the 50th ACM Technical Symposium on Computer Science Education},
}

@inproceedings{10.1145/3287324.3293708,
author = {Koning, Elizabeth and Adams, Joel C. and Hazlett, Christiaan D.},
title = {Visualizing Classic Synchronization Problems},
year = {2019},
booktitle = {Proceedings of the 50th ACM Technical Symposium on Computer Science Education},
}

@phdthesis{xie2008evaluating,
  title={Evaluating and refining diagrams that support the comprehension of concurrency and synchronization},
  author={Xie, Shaohua},
  year={2008},
  school={University of Georgia, Athens, GA, USA}
}

@article{kang2014visualization,
  title={Visualization tool for debugging data races in structured fork-join parallel programs},
  author={Kang, Myeong-Sin and Ha, Ok-Kyoon and Jun, Yong-Kee},
  journal={International Journal of Software Engineering and Its Applications},
  volume={8},
  number={4},
  pages={157--168},
  year={2014}
}

@article{2024-Estevez-ParallelEduLLM,
  title={Evaluation of LLM tools for feedback generation in a course on concurrent programming},
  author={Est{\'e}vez-Ayres, Iria and Callejo, Patricia and Hombrados-Herrera, Miguel {\'A}ngel and Alario-Hoyos, Carlos and Delgado Kloos, Carlos},
  journal={International Journal of Artificial Intelligence in Education},
  pages={1--17},
  year={2024},
  publisher={Springer}
}

@inproceedings{CodeAid,
author = {Kazemitabaar, Majeed and Ye, Runlong and Wang, Xiaoning and Henley, Austin Zachary and Denny, Paul and Craig, Michelle and Grossman, Tovi},
title = {CodeAid: Evaluating a Classroom Deployment of an LLM-based Programming Assistant that Balances Student and Educator Needs},
year = {2024},
booktitle = {Proceedings of the 2024 CHI Conference on Human Factors in Computing Systems},
}

@inproceedings{2025-Raihan-SIGCSE-LLMEduSurvey,
author = {Raihan, Nishat and Siddiq, Mohammed Latif and Santos, Joanna C.S. and Zampieri, Marcos},
title = {Large Language Models in Computer Science Education: A Systematic Literature Review},
year = {2025},
booktitle = {Proceedings of the 56th ACM Technical Symposium on Computer Science Education V. 1},
}

@article{2025-Yan-HSSC-LLMProgEdu,
  title={LLM-based collaborative programming: impact on students’ computational thinking and self-efficacy},
  author={Yan, Yi-Miao and Chen, Chuang-Qi and Hu, Yang-Bang and Ye, Xin-Dong},
  journal={Humanities and Social Sciences Communications},
  volume={12},
  number={1},
  pages={1--12},
  year={2025},
  publisher={Palgrave}
}

@ARTICLE{CEMR,
  author={Wan, Han and Luo, Hongzhen and Li, Mengying and Luo, Xiaoyan},
  journal={IEEE Transactions on Learning Technologies}, 
  title={Automated Program Repair for Introductory Programming Assignments}, 
  year={2024},
  volume={17},
  number={},
  pages={1705-1720},
}

@inproceedings{2024-Shen-SIGCSE,
author = {Shen, Yiyin and Ai, Xinyi and Soosai Raj, Adalbert Gerald and Leo John, Rogers Jeffrey and Syamkumar, Meenakshi},
title = {Implications of ChatGPT for Data Science Education},
year = {2024},
booktitle = {Proceedings of the 55th ACM Technical Symposium on Computer Science Education V. 1},
}

@article{GDB,
  title={Debugging with GDB},
  author={Stallman, Richard and Pesch, Roland and Shebs, Stan and others},
  journal={Free Software Foundation},
  volume={675},
  year={1988}
}

@misc{GPT-4o,
      title={GPT-4o System Card}, 
      author={OpenAI},
      year={2024},
      eprint={2410.21276},
      archivePrefix={arXiv},
      primaryClass={cs.CL},
      url={https://arxiv.org/abs/2410.21276}, 
}

@misc{Gemini, 
author = {{Google}},
title = {{Gemini}},
howpublished = {"\url{https://gemini.google.com/app}"},
year={2025}
}

@misc{Claude-Sonnet4, 
author = {{Anthropic}},
title = {{Claude Sonnet 4}},
howpublished = {"\url{https://www.anthropic.com/claude/sonnet}"},
year={2025}
}

\end{document}